# InfluenceTracker: Rating the impact of a Twitter account

Gerasimos Razis, Ioannis Anagnostopoulos

Computer Science and Biomedical Informatics Dpt., University of Thessaly

{razis, janag}@dib.uth.gr

**Abstract.** We describe a methodology of rating the influence of a Twitter account in this famous microblogging service. We then evaluate it over real accounts, under the belief that influence is not only a matter of quantity (amount of followers), but also a mixture of quality measures that reflect interaction, awareness, and visibility in the social sphere. The authors of this paper have created "InfluenceTracker", a publicly available website[1] where anyone can rate and compare the recent activity of any Twitter account.

## 1 Introduction

Microblogging is a form of Online Social Network (OSN) which attracts millions of users on daily basis. Twitter is one of these microblog services. Their users vary from citizens to political persons and from news agencies to huge multinational companies. Independent of the type of the user and of the degree of influence on other users, all share the same need; to spread their messages to as many users as possible.

The messages, which are regarded as pieces of information, can be spread in two ways, either directly or indirectly. A case of direct message is when a company reveals information about a new product to its Followers. When such a follower decides to share it among his or her own followers, then that is a case of indirect information dissemination.

In this paper, we propose a methodology for calculating the importance and the influence of a user in Twitter, as well as a framework, which describes the maximization of diffusion of information in such a network.

The remainder of this paper is organized as follows. In the next section, we provide an overview over the related work on discovering influential users in OSNs. Then, in Section 3 we describe the proposed methodology and the basic steps of the framework we use. In Section 4 real case scenarios are presented, in order to clearly show how we calculate the dissemination of information in Twittersphere, while in parallel we present the results along their assessment. Finally, Section 5 provides the conclusions of our work by summarizing the derived outcomes, while providing considerations on our future directions.

---

[1] http://www.influencetracker.com

## 2      Related Work

The calculation of the impact a user has on social networks, as well as the discovery of influencers in them is not new topic. It covers a wide range of sciences, ranged from sociology to viral marketing and from oral interactions to Online Social Networks (OSNs). In the related literature the term "influence" has several meanings and it is differently considered most of the times.

Romero et al. (Romero et al., 2011) utilize a large number of tweets containing at least one URL, their authors and their followers in order to calculate how influential or passive, in terms of activity, the Twitter users are. The produced influence metric depends on the "Follower-Following" relations of the users as well as their retweeting behavior. The authors state that the number of followers a user has is a relatively weak predictor of the maximum number of views a URL can achieve. As our work has shown (Section 3.2), through the retweet functionality information can be diffused to audience not targeted.

The study of Boyd et al. (Boyd et al., 2010) regarding several issues on retweets, has shown that this functionality along with the dissemination of information can be characterized as a conversational infrastructure. According to the authors, existence of conversation is regarded either when during a retweet some new information is added to the initial message, or when a single tweet is retweeted multiple times. The latter is interpreted by the authors as a statement of the users that they are either viewing the tweets or trying to invite new users into the conversation.

Cha et al. (Cha et al., 2010) introduce for each Twitter user three types of influence, namely "Indegree" (number of followers), "Retweet" (number of user generated tweets that have been reweeted) and "Mention" influence (number of times user is mentioned in other users' tweets). A necessary condition for the computation of these influence types is the creation of at least ten tweets per user. Those three types can also be calculated against a specific set of tweets which belong to a particular topic. The authors claim that "Retweet" and "Mention" influence correlate well with each other, while the "Indgree" does not. Therefore they suggest that users with high such influence type are not necessarily influential.

Another topic oriented study on the calculation of influence in OSNs is presented by Weng et al. (Weng et al. 2010). The authors propose an algorithm which takes into consideration both the topical similarity between users and their link structure. It is claimed that due to homophily, which is the tendency of individuals to associate and bond with others who have similar interests, most of the "Follower-Following" relations appear. This work also suggests that the active users are not necessarily influential.

The authors in (Anagnostopoulos et al., 2008) also directly associate social influence with homophily, as it plays an important role that may induce the directly related users to behave in a similar way. However, their proposed techniques have shown that despite the significant social correlation in the tagging behavior on Flickr, this correlation cannot be attributed to social influence.

Another approach which defines influence in terms of copying what the directly related do is presented by Goyal et al., (Goyal et al., 2010). In this work, the authors propose an influenceability score, which represents how easily a user is influenced by others or by external events. It is built on the hypothesis that a very active user performs actions without getting influenced by anyone. That type of users are regarded as responsible for the overall information dissemination in the network

All the related studies have shown that the most active users or those with the most followers are not necessarily the most influential. This fact has also been spotted by our work. Contrary to the aforementioned studies, for the calculation of our Influence metric we neither set a lower threshold on the number of the user-generated tweets, nor we utilize only a subset of these tweets which fulfill certain criteria (i.e. those which contain URL). All Twitter accounts are eligible for their calculation.

## 3 Methodology

As already mentioned, the contribution of this paper is twofold. Firstly, a methodology is proposed for calculating the importance and the influence of a user in Twitter. Secondly, a framework is described regarding its evaluation.

### 3.1 Calculation of user's Importance and Influence

Twitter users form a Social Network. If depicted in a graph, they would be represented by nodes. The edges that connect these nodes are the relations of "Follower-Following", introduced by Twitter. Obviously, some users are more influential than others. The methodology of calculating the importance and influence that a user has in an OSN is presented here.

That measurement should not depend merely on the number of "Followers" of a user, even if that number is big enough and the user's tweets are received by a large number of other users (followers). In case that the number of "Following" is larger, then the user could be characterized as a "passive" one. That type of users are regarded as those who are keener on viewing or being informed through tweets rather than composing new ones. Therefore, a suitable factor is the ratio of "Followers to Following" (FtF ratio).

But this ratio is also not sufficient. Another important factor is the tweets creation rate (TCR). For example, let us see the case where two users have nearly the same FtF ratio. Obviously the user with the higher TCR has more impact on the Network. In our methodology, in order to calculate that rate, we process the latest 100 tweets of the user according to the Twitter API. That leads to the TCR, and consequently the

Influence Metric, being dynamic as it depends on the most recent users' activity in Twitter.

The proposed Influence Metric depends on all of the aforementioned characteristics of the examined user, as defined in Equation 1. The FtF ratio is placed inside a base-10 log for avoiding outlier values. Moreover, the ratio is added by 1 so as to avoid the metric being equal to 0 in cases that the value of "Followers" is equal to "Following".

$$\text{Influence Metric} = \frac{\text{tweets}_k}{\text{Days}_{\text{since } k_{\text{th}} \text{tweet}}} * \text{OOM}(\text{Followers}) * \log_{10}\left(\frac{\text{Followers}}{\text{Following}} + 1\right),$$

$$\text{where OOM: Order Of Magnitude} \quad (1)$$

Each Tweet is associated with several other kinds of information. Two of them are the "ReTweets" and "Favorites" counts which represent how many times a Tweet has been retweeted or marked as favorite by other users. In our methodology, we utilize these counts in order to calculate the h-index of retweets and favorites counts over the last 100 tweets of an examined account. The aim of these measurements is to describe both the likeability and impact of the tweets of a Twitter user. These indexes are based on the established h-index measurement and are called as "ReTweet h-index - Last 100 Tweets" and "Favorite h-index - Last 100 Tweets".

Consequently, a Twitter account has "ReTweet h-index - Last 100 Tweets" h, if h over the last Nt tweets have at least h retweets each, and the remaining (Nt - h) of these tweets have no more than h retweets each (max. Nt=100).

The "ReTweet h-index - Daily" and the "Favorite h-index - Daily" are two similar metrics which represent the estimated daily value of "ReTweet and Favorite h-index" during the lifespan of a Twitter account over the last Nt tweets.

These h-index values are separately calculated and presented in influencetracker.com web site. However, we are currently working towards incorporating them in Equation 1, and more specifically on the evaluation of their impact over the proposed Influence Metric.

### 3.2 Information Diffusion/Tweet Transmission

An important functionality offered by Twitter is the "ReTweet". It allows users to repost a received tweet to their Followers. This results in viewing the tweets of users who are not being directly followed. That fact leads in the diffusion of information to users not targeted (to the followers of their followers). The same process can be repeatedly take place by the new viewers of the message and so on.

The most important factor which affects the transmission of the tweets is the followers' probability of retweeting. The higher this value is, the higher the probability of transmitting tweets to other users, initially not targeted by the source. Another dependency of the transmission of the tweets is the followers' TCR. The value of this rate includes both the users' generated tweets, as well as their retweets. The final dependency of that measurement is the "TCR of Follower to TCR of User" ratio.

Increased values of that ratio lead in bigger flow of tweets between these Twitter users.

The Tweet Transmission measurement depends on all of the aforementioned characteristics of the directly related users and it is defined in Equation 2.

$$\text{Tweet Transmission} = \frac{TCR_{n+1}}{TCR_n} * RT_{n+1}, \text{where } n \geq 0, n \in Z, 0 \leq RT \leq 1 \qquad (2)$$

### 3.3 Proposed evaluation framework

In order to evaluate the above metrics we employ the evaluation framework illustrated at Figure 1.

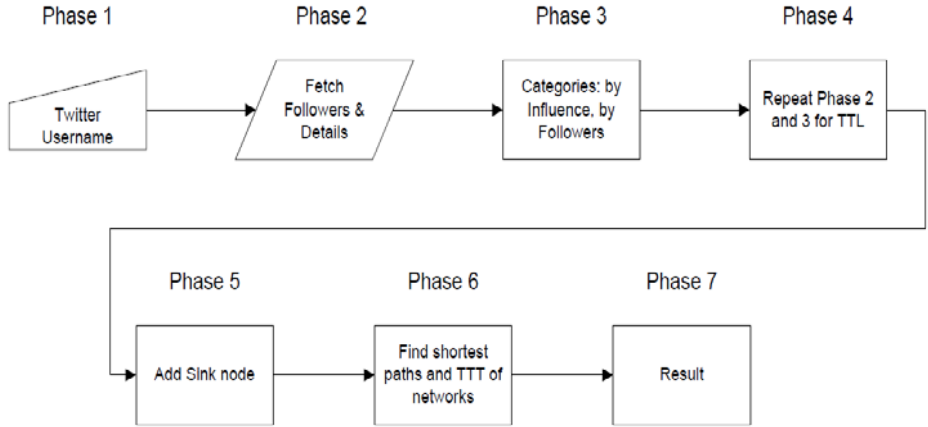

Figure 1: The seven phases of the proposed framework

The framework is split into seven Phases, presented in Figure 1. During the 1st Phase of the process, the Twitter user under examination is selected. In Phase 2, we fetch a large number of Followers ($N_f$) and their Twitter-related characteristics. These are necessary in order to calculate their Influence Metric measurement (Equation 1).

In Phase 3, all of these Followers are placed in two categories. The first one is classified by the value of our Influence Metric, while the second one by the absolute amount of followers each follower has. Both of these categories are sorted in descending format. After that, we select the top-k followers of these two categories.

Similarly, for these top-k users, Phases 2 and 3 are repeated. This process is continued until a specified distance threshold (layers) between Twitter users is reached (Phase 4). In computer networks, this distance is expressed by the Time-To-Live value (TTL) and corresponds to the amount of hops between different nodes a transmitted packet can perform before being rejected by the network. For the purposes of this work this threshold is set equal to 3.

The users, their followers, as well as their relations and characteristics are modeled as a separate network. Nodes depict users, while edges depict their relations containing specific attributes.

As a result of that process, two structures of the initial user and the Followers of Followers are created. The one depicts the top-k users by Influence, while the other one the top-k users by Followers. An example of such graph is presented in Figure 2.

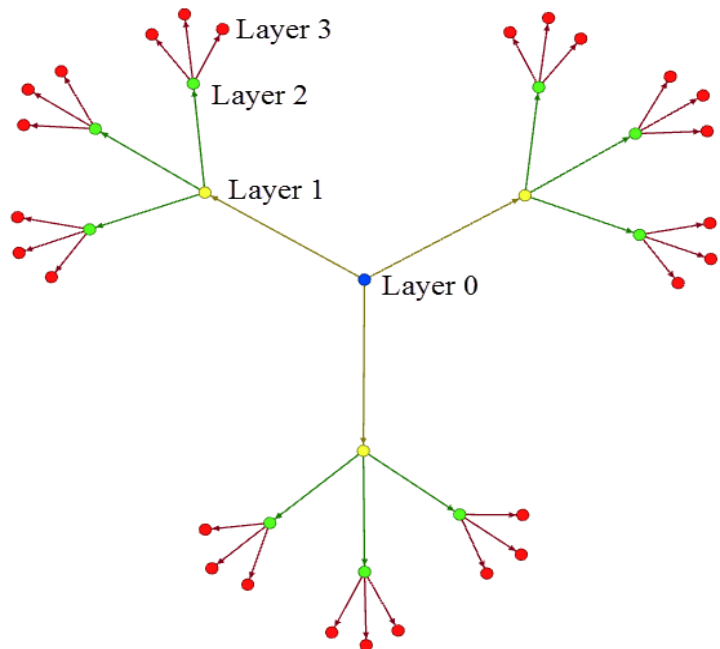

Figure 2: A 3-layered structure graph of the initial user and the top-3 Followers of Followers

In Figure 2 a 3-layered structure graph is displayed. The blue node represents the initial examined user. The user is connected with the yellow nodes, which stand for the top-3 followers either by the Influence Metric or by the amount of their followers (1st layer of distance). The process is iteratively continued with these nodes. The green and red nodes, 2nd and 3rd layer respectively, represent the followers of previously examined followers and so on. We should note here, that a node can be connected with others, independent if they belong to the same layer or not.

During the 5th Phase each of the two generated networks is terminated an ending node (sink) is added. This node is connected with all the users-followers of the last layer. These are the red nodes of Figure 2 which belong to the 3rd layer. That results in a fixed starting and ending point of the network. Figure 3 presents the network illustrated in Figure 2 including the sink node (black node in the center).

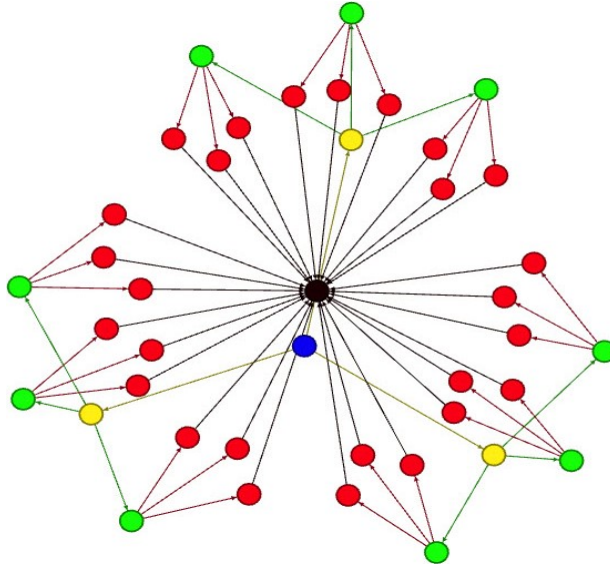

Figure 3: A 3-layered structure graph with a sink node

When all previous phases are completed, the sixth and final Phase is initiated. Its purpose is to discover all the paths, starting from the initial examined user (blue node) and ending to the sink (black node), and consist of exactly 4 steps. The number of 4 steps is very important, since it is the necessary and sufficient condition in order to find all the shortest paths between the initial user and the sink, which contain exactly one node belonging to every layer.

Furthermore, that number of steps ensures us that any possible loops will be avoided during the traversal of the networks from the user to the sink. A possible case of loop is when the examined Twitter user appears as a follower of another user. In such a case the initial user could also appear at the first (as a yellow node) or the second layer (as a green node).

The Tweet Transmission (TT) value, presented by Equation 2, is calculated for each layer of every shortest path. Then the TT value of the shortest path for all layers is calculated. This process is repeated until the TT values of all shortest paths of the two networks are computed. The network with the higher Total Tweet Transmission (TTT) value is considered the one with the higher disseminated information.

Figure 4 displays an example of a path derived from the network illustrated in Figure 3. This path consists of four edges (User→A, A→B, B→C, and C→Sink). For this path, three TT values are calculated, namely from User to A ($TT_{User \rightarrow A}$), from A to B ($TT_{A \rightarrow B}$), and finally from B to C ($TT_{B \rightarrow C}$). The total TT value of the whole path is the multiplication of these three calculated values and is assigned to the sink node. Figure 4 depicts the calculation of $TT_{A \rightarrow B}$.

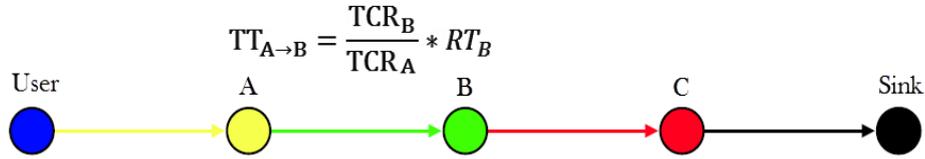

Figure 4: Calculation of a TT value

## 4 Evaluation results

In this section, we will present and analyze the results and the evaluation concerning the calculation of the importance and influence of a user in an OSN, and of the framework regarding the maximization of diffusion of information. As case study, we evaluate six real Twitter accounts. Three belong to political persons (@AdonisGeorgiadi, @IliasKasidiaris, and @PanosKammenos), one belongs to the Hellenic Fire Brigade (@Pyrosvestiki), and the rest belong to a Greek news media channel (@SkaiGr) and the international information network of activists and hacktivists named Anonymous (@YourAnonNews).

The experiments took place between 14/12/2013 and 31/1/2014. For each user four separate samplings were made, during which the number of the users' followers and their top-k users were gradually increased. The distance threshold, which defines the number of layers as described in Section 3.3, was set equal to 3.

### 4.1 Users' Influence

In this section, we present the Influence Metric measurements in respect to the examined Twitter accounts. We also provide the sampling date, the value of the Influence Metric, as well as other metrics are depicted in Table 1.

| ID | Username | Date | Influence | TCR | Followers | Following |
|---|---|---|---|---|---|---|
| AG1 | @AdonisGeorgiadi | 14/12/2013 | 126,857.416 | 15.50 | 33,410 | 3,574 |
| AG2 | @AdonisGeorgiadi | 18/12/2013 | 112,929.569 | 11.11 | 33,566 | 3,576 |
| AG3 | @AdonisGeorgiadi | 29/12/2013 | 511,537.359 | 50.00 | 34,164 | 3,579 |
| AG4 | @AdonisGeorgiadi | 16/01/2014 | 148,166.219 | 14.29 | 35,430 | 3,584 |
| IK1 | @IliasKasidiaris | 16/12/2013 | 26,686.871 | 1.11 | 14,148 | 56 |
| IK2 | @IliasKasidiaris | 19/12/2013 | 26,927.978 | 1.12 | 14,150 | 56 |
| IK3 | @IliasKasidiaris | 26/12/2013 | 25,492.531 | 1.06 | 14,172 | 56 |
| IK4 | @IliasKasidiaris | 26/01/2014 | 23,840.975 | 0.99 | 14,278 | 56 |
| PK1 | @PanosKammenos | 17/12/2013 | 63,708.939 | 3.33 | 33,889 | 419 |
| PK2 | @PanosKammenos | 21/12/2013 | 56,266.498 | 2.94 | 33,940 | 419 |
| PK3 | @PanosKammenos | 29/12/2013 | 46,724.779 | 2.44 | 34,029 | 419 |
| PK4 | @PanosKammenos | 12/01/2014 | 41,621.203 | 2.17 | 34,274 | 419 |
| P1 | @Pyrosvestiki | 01/01/2014 | 23,516.011 | 0.62 | 18,619 | 3 |
| P2 | @Pyrosvestiki | 30/01/2014 | 23,894.273 | 0.63 | 18,612 | 3 |
| P3 | @Pyrosvestiki | 31/01/2014 | 23,516.156 | 0.62 | 18,620 | 3 |
| P4 | @Pyrosvestiki | 31/01/2014 | 23,516.011 | 0.62 | 18,619 | 3 |
| SG1 | @SkaiGr | 17/12/2013 | 35,356,300.107 | 100.00 | 178,446 | 52 |
| SG2 | @SkaiGr | 21/12/2013 | 35,363,204.477 | 100.00 | 178,730 | 52 |
| SG3 | @SkaiGr | 31/12/2013 | 35,380,441.726 | 100.00 | 179,441 | 52 |
| SG4 | @SkaiGr | 01/01/2014 | 17,733,148.729 | 50.00 | 179,505 | 51 |
| YAN1 | @YourAnonNews | 18/12/2013 | 341,594,730.673 | 100.00 | 1,185,201 | 455 |
| YAN2 | @YourAnonNews | 23/12/2013 | 341,102,758.175 | 100.00 | 1,184,723 | 460 |
| YAN3 | @YourAnonNews | 27/12/2013 | 340,808,348.148 | 100.00 | 1,184,390 | 463 |
| YAN4 | @YourAnonNews | 24/01/2014 | 328,801,969.528 | 100.00 | 1,189,204 | 613 |

Table 1: The Influence Metric measurement and the Twitter related characteristics of the examined Twitter users

As we can see, the Influence Metric is directly dependant of the users' activity, which is measured by the TCR value. User's "@SkaiGr" Influence value during the first three samplings (SG1 to SG3) is approximately the same (nearly 35 Millions). However, during the fourth sampling (SG4) that value was almost the half. This was caused by the fact that the TCR value was dropped to half, despite that the "Followers to Following" ratio was slightly increased. In the case of user named "@YourAnonNews", during the first 3 samplings (YAN1 to YAN3) the Influence Metric value is nearly equal to 341 Millions. During the last sampling, YAN4, this value is dropped to approximately 329 Millions. This is explained due to the smaller value of "Followers to Following" ratio (the amount of following users increased during the period of the last sampling).

We should note here, that for the calculation of the Influence Metric, we consider the latest 100 users' tweets directly from the Twitter API. This enables the measurement to be dynamic and in accordance to the latest trend activity of the examined Twitter account.

## 4.2 Information Diffusion/Tweet Transmission

In this section, the TTT values of the created networks in respect of the examined users are presented. Table 2 is divided in four parts. Each part refers to the separate samplings mentioned above, while the related information per part is:

- the number of followers that is iteratively fetched,
- the number of top-k followers of the two generated categories, these are "by Influence Metric" and "by Followers", which are used for the creation of the respective layered networks,
- the distance threshold value that reflects the layers of the examined user's networks (TTL),
- the user who is the root of the two resulting networks (the six examined users),
- the TTT values of the two networks, according to the "by Influence Metric" and "by Followers", and finally
- the difference of the above TTT values for both generated networks.

In addition, the green-highlighted values in column "Difference" correspond to the cases where the TTT value is larger in the "By Influence" category, thus indicating that our approach manages to create a network of followers who are more influential in comparison to the network of category "By Followers". Red-highlighted values reflect to the opposite cases. As we can see the wider the examined networks are in terms of the top-k users and their followers up to third layer of difference, the more influential network of users we have.

As can be observed from Table 2, the TTT values of the two networks are escalated as both the numbers of the Followers and of the top-k users are also increased.

The results of the use cases used for the evaluation of the influence metric calculation, show that the number of followers a user has, is not solely sufficient to guarantee the maximum diffusion of information in Twitter (and practically to any similar OSN). This is because, these followers should not only be active Twitter users, but also have impact on the network. The latter is calculated by the Influence Metric value.

| Followers = 50, top-k users = 3, TTL = 3 |||| 
| User | By Influence | By Followers | Difference |
|---|---|---|---|
| @AdonisGeorgiadi | 2.174 | 12.933 | -10.759 |
| @IliasKasidiaris | 57.833 | 42.027 | 15.806 |
| @PanosKammenos | 22.527 | 30.074 | -7.547 |
| @SkaiGr | 1.016 | 0.465 | 0.551 |
| @YourAnonNews | 0.038 | 0.018 | 0.020 |
| @Pyrosvestiki | 0.496 | 0.864 | -0.368 |

| Followers = 100, top-k users = 5, TTL = 3 ||||
| User | By Influence | By Followers | Difference |
|---|---|---|---|
| @AdonisGeorgiadi | 12.733 | 8.632 | 4.102 |
| @IliasKasidiaris | 116.048 | 241.823 | -125.775 |
| @PanosKammenos | 134.417 | 30.997 | 103.420 |
| @SkaiGr | 3.462 | 0.302 | 3.160 |
| @YourAnonNews | 1.762 | 0.446 | 1.316 |
| @Pyrosvestiki | 210.442 | 85.437 | 125.005 |

| Followers = 180, top-k users = 7, TTL = 3 ||||
| User | By Influence | By Followers | Difference |
|---|---|---|---|
| @AdonisGeorgiadi | 45.831 | 20.038 | 25.794 |
| @IliasKasidiaris | 682.280 | 592.961 | 89.319 |
| @PanosKammenos | 723.534 | 373.959 | 349.575 |
| @SkaiGr | 4.773 | 3.172 | 1.600 |
| @YourAnonNews | 2.234 | 0.980 | 1.255 |
| @Pyrosvestiki | 909.388 | 263.730 | 645.658 |

| Followers = 360, top-k users = 7, TTL = 3 ||||
| User | By Influence | By Followers | Difference |
|---|---|---|---|
| @AdonisGeorgiadi | 50.686 | 15.503 | 35.183 |
| @IliasKasidiaris | 124.871 | 265.954 | -141.083 |
| @PanosKammenos | 549.347 | 108.909 | 440.438 |
| @SkaiGr | 3.768 | 2.628 | 1.141 |
| @YourAnonNews | 3.844 | 2.866 | 0.978 |
| @Pyrosvestiki | 917.656 | 533.136 | 384.520 |

Table 2: The tables containing the details of each sampling set

## 5      Conclusions and Future Work

In this paper, we proposed a methodology for calculating the importance and the influence of a Twitter account, as well as a methodology regarding the maximization of diffusion of information. For evaluation purposes a framework was applied. As a case study, we evaluated six real Twitter accounts. The experiments took place between 14/12/2013 and 31/1/2014 and for each account four separate samplings were made.

The results of the use cases show that the number of followers a user has, is not sufficient to guarantee the maximum diffusion of information in Twitter (and practically to any similar OSN). This is because, these followers should not only be active Twitter users, but also have impact on the network. The latter is calculated by our Influence Metric.

Ongoing research is performed on how the proposed Influence Metric can be improved. We are currently working towards incorporating the "ReTweet h-index - Daily" and the "Favorite h-index - Daily" in Equation 1 and more specifically on the evaluation of their impact over the proposed Influence Metric.